\documentclass[letterpaper, 10 pt, conference]{ieeeconf}  % Comment this line out if you need a4paper

\IEEEoverridecommandlockouts                              % This command is only needed if 
\usepackage{amsmath}
\usepackage{amssymb}                   
\usepackage{mathrsfs}

\usepackage{psfrag}
\usepackage[dvipsnames]{xcolor} % purple, violet, teal // colortbl to color tables
\usepackage{tabularx,colortbl}

\definecolor{listinggray}{gray}{0.9}
\definecolor{lbcolor}{rgb}{0.9,0.9,0.9}

\usepackage{tikz}
\usetikzlibrary{positioning}
\usetikzlibrary{arrows.meta}
\usetikzlibrary{decorations.pathmorphing}
\usepackage[detect-all]{siunitx}
\usetikzlibrary{backgrounds}

\newcommand\vi{\vspace{\baselineskip}}

\newcommand\be{\begin{equation}}
\newcommand\ee{\end{equation}}
\newcommand\ba{\begin{aligned}}
\newcommand\ea{\end{aligned}}
\newcommand\bma{\begin{bmatrix}}
\newcommand\ema{\end{bmatrix}}

\newcommand{\Rbb}{\mathbb{R}}
\newcommand{\Zbb}{\mathbb{Z}}

\newcommand{\rank}{\mathrm{rank}}

\newcommand{\bbm}[1]{\left[\begin{matrix} #1 \end{matrix}\right]}
\newcommand{\sbm}[1]{\left[\begin{smallmatrix} #1
   \end{smallmatrix}\right]}

\newcommand{\rfb}[1]{\mbox{\rm
   (\ref{#1})}\ifx\undefined\stillediting\else:\fbox{$#1$}\fi}

\newtheorem{theorem}{\it Theorem}
\newtheorem{lemma}{\it Lemma}
\newtheorem{definition}{\it Definition}

\title{\LARGE \bf
On the one-shot data-driven verification of dissipativity of LTI systems with general quadratic supply rate function 
}

\author{Tábitha E. Rosa$^{1}$ and Bayu Jayawardhana$^{1}$% <-this % stops a space
\thanks{*This work was supported by the STW project 15472 of the STW Smart Industry 2016 program}% <-this % stops a space
\thanks{$^{1}$Tábitha E. Rosa and Bayu Jayawardhana are with the ENgineering and TEchnology institute Groningen (ENTEG), 
        University of Groningen, Nijenborgh 4, 9747 AG Groningen, The Netherlands
        {\tt\small \{t.estevesrosa,b.jayawardhana\}@rug.nl }}%
}

\begin{document}

\maketitle
\thispagestyle{empty}
\pagestyle{empty}

%%%%%%%%%%%%%%%%%%%%%%%%%%%%%%%%%%%%%%%%%%%%%%%%%%%%%%%%%%%%%%%%%%%%%%%%%%%%%%%%
\begin{abstract}

Based on a one-shot input-output set of data from an LTI system, we present a verification method of dissipativity property based on a general quadratic supply-rate function. We show the applicability of our approach for identifying suitable general quadratic supply-rate function in two numerical examples, one regarding the estimation of $\mathcal{L}_2$-gains and one where we verify the dissipativity of a mass-spring-damper system.
\end{abstract}


 %%%%%%%%%%%%%%%%%%%%%%%%%%%%%%%%%%%%%%%%%%%%%%%%%%%%%%%%%%%%%%%%%%%%%%%%%%%%%%%%

\section{INTRODUCTION}
The use of model-based control design has dominated the landscape of control systems in the previous century. In this case, the (actuator, plant and sensor) systems dynamics are described as state equations or transfer functions based on the underlying first principle models and systems identification methods. Subsequently, they are used to design the controllers in order to meet a number of control specifications, including, stability and robustness of the closed-loop system. For the latter, the notion of dissipative systems has played a key-role in defining the concept of $L_2$-stability and $\mathcal{H}_\infty$~robust control~\cite{Doyle1989}.   

The rise of system-of-systems, where cyber-physical systems are interconnected with each other, has resulted in complex systems that are hard-to-model. While they can produce a large number of data through the network of sensors in the systems, the lack of computationally tractable model has limited the applicability of the big data for control design and for guaranteeing stability and robustness. Correspondingly, the data-driven input-output characterization of such complex systems, which can be suitable for control design purposes, has received a renewed interest in recent years. In this paper, we are interested in the particular data-driven characterization of LTI systems, namely, the dissipativity property which has been instrumental in the development of model-based $\mathcal{H}_\infty$ robust control design.  
%Lately, much has been discussed about {\em data-driven} systems, or, the also called {\em data-based} systems. Until only a few decades ago, the steps for obtaining a controlled system would comprise either the modelling of the system and then the design of its controllers, or implementations of controllers based on trial and errors and/or the experience of the control designer. With the advances of technology, it became easier to obtain and deal with data, and, consequently, to design controllers from this ample source of information, yielding researchers everywhere to invest their times on this subject. However, even though we can see that important steps have been taken towards this direction, we still lack some theoretical basis for analysing such systems. Important issues as, for instance, analysing {\em dissipativity} and {\em stability} properties of data-driven systems are still open problems.

The dissipative systems concept has been studied since the '70s with the seminal works of Jan Willems~\cite{Wil:72}, and Hill and Moylan~\cite{Hill1976,HM:80}. In these works, the dissipative systems can be described based on their input-output behaviours, as well as, on the state space realization. The study of dissipative systems was motivated by physical systems where energy functions can be defined for such systems that satisfy energy conservation laws. In other words, the rate change of the energy functions, the so-called storage functions, are upper bounded by the power or work done to the systems which are commonly referred to as the supply-rate functions. 

In systems theory, the specific structure of supply-rate functions can be used to determine the stability property of the dissipative systems. When the supply-rate function is given by the product of input and output, the corresponding dissipative systems are called passive systems. It includes the well-studied Euler-Lagrange and port-Hamiltonian systems where the input is given by the generalized forces and the output is given by the generalized velocity. For dissipative systems with supply-rate functions given by $-k\|y\|^2 + \|u\|^2$, where $k>0$, $u$ is the input and $y$ is the output, they are called $L_2$-stable systems with the $L_2$-gain of $k$. The associated dissipative inequality is also used in robust control design. A larger class of dissipative systems is defined by the supply-rate functions  $\sbm{y^T&u^T}\sbm{Q&S\\S^T&R}\sbm{y\\u}$ with symmetric matrices $Q$ and $R$, which are known as the $QSR$-dissipative systems \cite{Hill1976}. Other non-standard supply-rate functions include counterclockwise systems/negative imaginary systems with supply-rate function  $\langle\dot y,u\rangle$ \cite{Lanzon2008,Ouyang2014} and clockwise systems with supply-rate function $\langle y,\dot u\rangle$ \cite{Ouyang2013}. In fact, via behavioural framework, it has been shown that all linear systems can be characterized through a specific form of quadratic supply-rate functions that involves the input $u$, output $y$ and all its derivatives (up to the order of the systems) \cite{TW:97}. 

Correspondingly, we investigate in this paper the characterization of such general supply-rate function for discrete-time LTI systems, particularly, based only on one-shot of input-output data, e.g., a segment of any given input-output trajectory. %For discrete-time systems, the quadratic form involving the derivatives of $u$ and $y$ becomes that involving the time-difference of $u$ and $y$, respectively. 
Such data-driven characterization can be of practical use when we do not have a complete model/state-space knowledge. The availability of such information can further be used to determine the stability of feedback interconnection of complex cyber-physical systems, for instance.  

%were the pioneers in the dissipative theory, a theory that explain the dynamic of a system through the analysis of its input and output. Their main idea is that a system is dissipative by verifying an inequality that contains a {\em storage function} and a {\em supply rate}, being the storage function a function that evaluates the quantity of energy that is stored inside the system and the supply rate as the rate with which the energy flows into the system. Even though their theory is based on an input-output formulation, the methods to verify the dissipativity inequality were vastly extended and understood in terms of applicability to real-world applications only in the model-based domain. This phenomenon might be explained by the lack of interested on data-based approaches given that scarcity of technologies for allowing such approaches at that point in time. 

%With the increase of computational resources of the past decades, the interest in studying data-driven systems also increased. 
In recent literature on data-driven identification and control, the concept of {\em persistency of excitation} \cite{WRMDM:05}, which is also known as the {\em Willems' fundamental lemma}, plays an important role to obtain the set of behaviours and to subsequently use them for characterising various discrete-time systems properties.
Based on such concept, several data-driven approaches were proposed in the literature, where most of them refer to model identification and/or control, we refer interested readers to, for instance,~{\cite{dPT:19,BA:19,MR:08}.} 

A number of methods have been proposed recently to verify the dissipativity of data-driven {\em linear time invariant} (LTI) systems have been proposed, such as,  \cite{MM-MR:17,RBKA:19}. In \cite{MM-MR:17}, the authors propose a method in the behavioural framework to verify the dissipativity of an LTI system using a quadratic differential supply function that was investigated in \cite{TW:97}. The approach in \cite{MM-MR:17} results in a dissipative verification algorithm  based on solving a non-convex indefinite quadratic program.  %Here, the QSR-dissipativity coincides with that studied in %Normally, methods for verifying dissipativity properties make use of the so called $QSR$-dissipativity that is investigated in 
%\cite{Hill1976,HM:80}. %, a method where the supply function is quadratic with respect to the input and output in the present time. Thus, the work of \cite{MM-MR:17} is a more general approach and allows the control designer to have more flexibility in terms of how to analyse the system. 
In \cite{RBKA:19}, the authors recast the problem in %extend the work in 
\cite{MM-MR:17} into a convex problem using the standard $QSR$-dissipative supply function. The approach has been shown to work well for several practical applications in \cite{RBKA:19}. % to the usual $QSR$-dissipativity, which already works well for several practical situations. However, in the case of data-driven systems, utilizing only the $QSR$ approach can lead to conservative results, which happens, for instance, in the case of delay systems and in cases where the storage function produces a supply function that requires information present in more time samples than only the present time.

%{\color{blue} Bayu will continue on 19 Nov evening}\\ 

Inspired by \cite{RBKA:19}, we extend the work of \cite{RBKA:19} by considering a general quadratic supply-rate function that can capture the behaviour of all linear systems \`a la \cite{TW:97} in the continuous-time case. More precisely, we propose a data-driven method based on one-shot input-output data for testing the dissipativity with respect to such general quadratic supply-rate functions. It provides us a mean to identify the admissible form of general quadratic supply-rate function which can potentially be used to help finding an admissible storage function, as well as, to determine the stability of interconnected data-driven systems, beyond the standard passivity interconnection. 
%In this paper, we introduce an approach for verifying the dissipativity of a system using a quadratic supply function that allows time-differences using only one shot of data. We consider a discrete-time LTI system for our analysis. We use the concepts of $L$- and $(L-\nu)$-dissipativity for obtaining our main results, and we also provide reasoning on how to choose the the parameters necessary for the verifying them. 
We illustrate our method using two different examples. In the first one, we consider stable LTI systems (with bounded $\mathcal L_2$-gain) where we validate and compare our approach in verifying the $\mathcal L_2$-stability of the systems with respect to that presented in \cite{RBKA:19}. In the second one, we consider a typical mass-spring-damper system and verify if our conditions hold knowing that the system is already dissipative with respect to a known supply function.  

\vi

{\em Notation:} 
The set of vectors (matrices) of order $n$ ($n\times m$) with real entries is represented by $\Rbb^n$ ($\Rbb^{n\times m}$), for integer entries the equivalent is represented by $\Zbb^n$ ($\Zbb^{n\times m}$). Similar notation is applied to denote %the dimensions of 
a vector (matrix) with zero and ones by % and identity forms, 
$0^n$ and $1^n$ (or $0^{n\times m}$ and $1^{n\times m}$), respectively. The $n\times n$ identity matrix is denoted by $I^n$. Additionally, we use a subscript $+$ or $-$ to denote sets with only positive or negative, respectively, % A set that contains positive (negatives) entries is defined as, 
for instance, $\Zbb_+$ ($\Zbb_-$) that denotes a set of positive (negative) integers. 
For matrices or vectors, the symbol $^\top$ denotes the transpose. A positive (or negative) symmetric matrix $P$ is %positive (or negative) definite is represented 
denoted by $P\succ0$ (or $P\prec0$).
The space of discrete signals that are square summable is defined by $\ell_2(\Rbb^\bullet)$.
Given $e\in \ell_2(\Rbb^\bullet)$, we denote $\{e\}_{i}^{j}=\{e(i),\ldots,e(j)\}$, we define its stacked vector by
\be 
    e_{[i,j]}=\bma e(i)^\top&e(i+1)^\top&\cdots&e(j)^\top\ema^\top .
\ee
Throughout the paper, we use them interchangeably whenever it is clear from the context.

A Hankel matrix with $L\in\Zbb_+$ block rows of a finite sequence $e_{[0,T-1]}$ is given by
\be \small
    H_{L } ({e}_{[0,T-1]}) = \bma e(0) & e(1) &\cdots&e(T-L)\\
    e(1) & e(2) &\cdots&e(T-L+1)\\
    \vdots & \vdots & \ddots &\vdots\\
    e(L-1) & e(L-2) &\cdots&e(T-1)\ema.
\ee

\begin{definition}[\cite{vWdPCT:20,WRMDM:05}]\label{def:persist_simple}
    A measured trajectory $e_{[0,T-1]}$, $e:\Zbb \rightarrow \Rbb^n$ is {\em persistently exciting} of order $L$ if $\rank (H_L(e_{[0,T-1]}))=nL$.
\end{definition}
% From this last definition and also the one of a Hankel matrix, we have that $T\geq (n+1)L-1$.

\begin{lemma}[Finsler's Lemma \cite{dOS:01}]
	\label{lema:finsler}
    If there exist $w\in\Rbb^{n}$, $Q\in\Rbb^{n\times n}$, $B\in\Rbb^{m\times n}$ with $\rank(B)<n$, and $B^\perp$ is a basis for the null space of $B$, that is, $BB^\perp=0$, then all the following conditions are equivalent
	\begin{enumerate}
		\item $w^\top Q w< 0,~\forall w \neq 0~:~B w=0$,
		\item ${B^\perp }^\top QB^\perp <0 $,
		\item $\exists \mu\in\Rbb~:~Q-\mu B^\top B<0$,
		\item\label{item:finsler} $\exists\mathcal{X}\in\Rbb^{n\times m}~:~Q+XB+B^\top X^\top <0$.
	\end{enumerate}
\end{lemma}

\section{Problem Formulation}
Consider the following causal discrete-time linear time-invariant (LTI) system 
\be \label{eq:sys_main}
    \Sigma : \left\{\begin{aligned} {x}(k+1)=&~Ax(k)+Bu(k),\\
    y(k)=&~Cx(k) +Du(k),\\
    x(0)=&~x_{0}\end{aligned}\right.
\ee
where $x(k)\in\Rbb^{n}$ is the state vector, $u(k)\in\Rbb^m$ is the control input and $y(k)\in\Rbb^p$ is the output. 
We assume that the state space matrices are unknown, however, we do have access to the input and output information for all $k=0,\ldots,T_f$, where $T_f$ is any arbitrary given time. Following the works of Willems \cite{Wil:72} and Hill and Moylan \cite{HM:80}, system \eqref{eq:sys_main} is assumed to be {\em dissipative} with respect to a supply rate $w(y(k),u(k))$ as defined below. The manifest variable of $\Sigma$ is denoted by $z(k) = \bbm{y(k)^\top & u(k)^\top}^\top$. 

\begin{definition}[Dissipativity \cite{BLME:07}]\label{def:dissip}
    System \eqref{eq:sys_main} is said to be dissipative with respect to a supply rate $w(y(k),u(k))$ if there exists a storage function $V:\Rbb^n\rightarrow\Rbb_+$ with $V(0)=0$ such that
    \be \label{eq:lyapmb} 
        V(x(k)) -  V(x(0)) \leq \sum_{i=0}^{k-1}w(y({i}),u({i}))
    \ee
    or equivalently,
    \be \label{eq:lyapmb2} 
        V(x(k+1)) -  V(x(k)) \leq w(y({k}),u({k}))
    \ee
    holds along all possible  trajectories of \eqref{eq:sys_main} for all $k\geq 0$,   %starting at $x_{0}$, 
    $x_{0}\in\Rbb^n$ and $u\in \ell_2(\Rbb^m)$.  
\end{definition}

In the context of model-based, we have several methods that can be used to both verify if the system is dissipative with respect to a certain supply rate and to find such supply rate~\cite{HM:80,BLME:07,KA:10}. One particular approach that has been extensively studied in literature is the $QSR$-dissipativity \cite{HM:80}. %, which is presented in the following definition.
%\begin{definition}[$QSR$-dissipativity \cite{HM:80}]\label{def:QSR}
     System \eqref{eq:sys_main} is said to be {\em $QSR$-dissipative} with respect to a  supply rate
    \be \label{eq:supplyfunction}
        \begin{aligned} 
        w(u(k),y(k)) %=&~ \bma y(k) \\u(k) \ema^\top \Phi\bma y(k) \\u(k) \ema\\
        =&~ \bma y(k) \\u(k) \ema^\top \underbrace{\bma Q &S \\S^\top &R\ema}_{=:\Phi}\bma y(k) \\u(k) \ema, 
        \end{aligned}
    \ee
    where $Q=Q^\top \in\Rbb^{p\times p}$, $R=R^\top \in\Rbb^{m\times m}$ and $S\in\Rbb^{p\times m}$,
    if there exists a {\em storage function} $V:\Rbb^n\rightarrow\Rbb_+$ with $V(0)=0$ such that \eqref{eq:lyapmb} and \eqref{eq:lyapmb2} holds along all possible trajectories of $\Sigma$ for all $k\geq 0$, $x_{0}\in\Rbb^n$ and $u\in \ell_2(\Rbb^m)$. %starting at $x_{0},~k\geq0$ and $u{(k)}$. 
%\end{definition}

As we are interested in investigating the dissipativity of data-driven linear systems, which are defined solely based on the available measurement data $y$ and $u$, the above definition is no longer suitable as we may not have access to the state variables. As introduced in \cite{MM-MR:17}, another approach to verify the $QSR$-dissipativity of a data-driven linear system satisfying~\eqref{eq:sys_main}  %from a given input and output data of $u$ and $y$ %\eqref{eq:sys_main} 
%is $QSR$-dissipative 
is simply by evaluating the following inequality 
\be\label{eq:QSRdissip} 
    \sum_{k=0}^\infty w(u(k),y(k)) \geq0,
\ee
where the supply function is given by \eqref{eq:supplyfunction} and its initial state is taken to be zero $x_{0}=0$. When non-zero initial conditions are considered, then the lower-bound in \eqref{eq:QSRdissip} will be a negative-definite function of $x_0$ \cite{Wil:72}. 
Furthermore, the data-driven system \eqref{eq:sys_main} is said to be {\em $L$-$QSR$-dissipative} if 
\be\label{eq:LQSRdissip} 
    \sum_{k=0}^{L-1} w(u(k),y(k)) \geq0
\ee
holds for all trajectories ($u_{[0,L-1]},y_{[0,L-1]}$) with $w$ as in \eqref{eq:supplyfunction} and $x_{0}=0$. This last approach has also been explored in \cite{RBKA:19}, in the state-space framework, where the authors introduce an approach to verify the $L$-$QSR$-dissipativity properties of a data-driven system as in \eqref{eq:sys_main} using only one batch of data. Additionally, the paper \cite{RBKA:19} introduces the notion of {\em ($L-\nu$)-$QSR$-dissipativity} properties which we explore throughout our main results. For the sake of completeness, let us define this as follows. For a given $n\leq \nu<L$, the system \eqref{eq:sys_main} is said to be {\em $(L-\nu)$-$QSR$-dissipative} if 
\be\label{eq:LnuQSRdissip} 
    \sum_{k=0}^{L-1-\nu} w(u(k),y(k)) \geq0
\ee
holds for all trajectories ($u_{[0,L-1-\nu]},y_{[0,L-1-\nu]}$) with $w$ as in~\eqref{eq:supplyfunction} and $x_{0}=0$. 

%{\color{red} Define the $L-QSR$ and $(L-\nu)-QSR$ dissipativity here.}

An important assumption for satisfying \eqref{eq:LQSRdissip} is that given the input of the measured trajectories being persistently exciting, we can obtain other admissible trajectories of the system using a single shot of data \cite{vWdPCT:20,WRMDM:05,dPT:19,BA:19,MM-MR:17,RBKA:19}.
This concept is presented formally in the following lemma.
\begin{lemma}\label{lem:persist}
    Consider a vector $z(k) = \bma y(k)^\top & u(k)^\top \ema^\top $ and a measured trajectory $z_{[0,T-1]}$ with $u(k)$ being persistently exciting of order $L+n$.  Then, a set of data given by $\bar{z}_{[0,L-1]}$ is a trajectory of $\Sigma$ if and only if there exists a vector $\alpha\in\Rbb^{T-L+1}$ such that
    \be \label{eq:persist} 
        H_{L}(z_{[0,T-1]})\alpha = \bar{z}_{[0,L-1]} .
    \ee
\end{lemma}

As presented in \cite{BA:19}, we have that \eqref{eq:persist} is equivalent to 
\[ 
    \bar{z}_{[0,L-1]} = \sum_{i=0}^{T-L}\alpha_i {z}_{[i,L-1+i]}.
\]
It means that if the input of the system is persistently excited, we can obtain a complete set of trajectories through a linear combination of the initial set of measured data considering time shifts. The proof of Lemma~\ref{lem:persist} has been explored in several works in both contexts of state-space and behavioral systems \cite{vWdPCT:20,WRMDM:05,MM-MR:17,dPT:19}.

Regarding the supply function, in this paper, we are interested to study a case similar to \cite{MM-MR:17}, where we assume a supply-rate function that include the time differences of the measured data\footnote{Note that in the continuous-time domain, this is equivalent to consider the derivatives of the inputs and outputs in the supply function.}. Particularly, we consider the following general supply-rate function 
\be  \label{eq:supplyfunctionderi} 
\begin{aligned}
    w(u(k),y(k)) =&~ \sum_{i,j=0}^N\bma y(k+i) \\u(k+i) \ema^\top\Phi_{ij}\bma y(k+j) \\u(k+j) \ema\\
     =&~\sum_{i,j=0}^N z(k+i)^\top\Phi_{ij}z(k+j)
\end{aligned}
\ee
where each $\Phi_{ij}$ is a $QSR$ matrix as given before in \eqref{eq:supplyfunction} that describes the relation between a pair of measurement data $z(k+i)$ and $z(k+j)$. % time difference, 
Furthermore, we consider
\[
    \Phi_N = \bma \Phi_{00} & \cdots & \Phi_{0N}\\
    \vdots  & \ddots & \vdots\\
    \Phi_{N0} & \cdots & \Phi_{NN}
    \ema,
\]
with $\Phi_{ji}=\Phi_{ij}^\top$ for all $i,j=\{0,\ldots,N\}$. In the following, $\Sigma$~is said to be $N$-$QSR$ dissipative if the dissipative inequality~\eqref{eq:QSRdissip} holds with the supply-rate be given by \eqref{eq:supplyfunctionderi} for the given $N$. In the same manner, it is $(L,N)$-$QSR$ (or $(L,\nu,N)$-$QSR$) dissipative if \eqref{eq:LQSRdissip} (or correspondingly,~\eqref{eq:LnuQSRdissip}) holds with the supply-rate be given by \eqref{eq:supplyfunctionderi} for the given $L$, $N$ (and $\nu$), respectively.

%%%%%%%%%%%%%%%%%%%%%%%%%%%%%%%%%%%%%%%%%%%%%%%%%%%%%%%%%%%%%%%%%%%%%%%%%%%%%%%%%%%
% MAIN RESULTS
%%%%%%%%%%%%%%%%%%%%%%%%%%%%%%%%%%%%%%%%%%%%%%%%%%%%%%%%%%%%%%%%%%%%%%%%%%%%%%%%%%%

\section{Main results}

In this section, we present our main results. In the following theorem, we introduce a method to verify the $(L,N)$- and $(L,\nu,N)$-$QSR$-dissipativity of an LTI system with respect to a general $QSR$ form as in \eqref{eq:supplyfunctionderi}. %, which can be seen in the following theorem, proceeded by additional comments and propositions. 

%%%%%%%%%%%%%%%%%%%%%%%%%%%%%%%%%%%%%%%%%%%%%%%
%              THEOREM
%%%%%%%%%%%%%%%%%%%%%%%%%%%%%%%%%%%%%%%%%%%%%%%

\begin{theorem}\label{theo:LQSR}
Let $z(k) = \bma y(k)^\top & u(k)^\top \ema^\top$ and suppose that $z_{[0,T+N-1]}$ is a given measured trajectory of $\Sigma$ in  \eqref{eq:sys_main}. Then the following statements hold.
\begin{itemize}
\item[i)] System \eqref{eq:sys_main} is $(L,\nu,N)$-$QSR$ dissipative if $u_{[0,T+N-1]}$ is persistently exciting of order $L+N+n$~and
\be \label{eq:theo}
   {U_\perp}^\top H_{L}(\mathcal{Z}_{[0,T-1]})^\top \Phi_L H_{L}(\mathcal{Z}_{[0,T-1]}){U_\perp}\succeq 0
\ee 
where
%\begin{equation}
$\Phi_L = I_L \otimes \Phi_N$,  
   % \Phi =&~ \bma 
   % \Phi_{00}  & \Phi_{10}  & \cdots & \Phi_{N0} \\
   % \Phi_{01}  & \Phi_{11}  & \cdots & \vdots    \\
  %  \vdots     & \vdots     & \ddots & \vdots    \\
   % \Phi_{0N}  & \Phi_{1N}  & \cdots & \Phi_{NN} 
  %  \ema .
%\end{equation}
%$\{\mathcal{Z}(k)\}_{k=0}^{T-1}$ than this
 the data contained in $z_{[0,T+N-1]}$ is rearranged in the form of $\mathcal{Z}_{[0,T-1]}$ with $\mathcal{Z}(k) = \bma z(k)^\top & \cdots& z(k+N)^\top \ema^\top $,  and considering $ U_\perp=({U} H_L(\mathcal{Z}_{[0,T-1]}))^\perp$ where
\be\label{eq:U_def} 
\begin{split}
    U = \bma  U_{\text{\rm aux}}  & 0^{(m+p)\nu\times (m+p)(L-\nu)(N+1)}\ema,\\
    U_{\text{\rm aux}} = I^\nu\otimes  \begin{pmatrix} I^{m+p} & 0^{(m+p)\times (m+p)N} \end{pmatrix} ,
    \end{split}
%U = \bbm{I^{(m+p)\nu} & 0^{(m+p)\nu\times (m+p)(L(N+1)-\nu)}}
\ee
for some $\nu<L$. %{\color{purple} the old $U$ is the correct one. If we consider your new one, it means that it would be a big identity and zeros next to it, but that's not what we need.. we need to a Kronecker product such that we repeat $\nu$ times a matrix that is an identity and zeros.}
\item[ii)] Additionally, if the inequality \eqref{eq:theo} holds for any $n\leq \nu<L$, then system~\eqref{eq:sys_main} is $(L,N)$-$QSR$-dissipative.
\end{itemize}
\end{theorem}

%%%%%%%%%%%%%%%%%%%%%%%%%%%%%%%%%%%%%%%%%%%%%%%
%               PROOF
%%%%%%%%%%%%%%%%%%%%%%%%%%%%%%%%%%%%%%%%%%%%%%%

\textit{Proof:} 
Suppose that the hypotheses in Theorem \ref{theo:LQSR}i) hold where we have $u_{[0,T+N-1]}$ being persistently exciting of order $L+N+n$ and there exists $\nu<L$ such that \eqref{eq:theo} holds. From Definition~\ref{def:persist_simple} and from Lemma \ref{lem:persist}, we know that if $u$ is persistently exciting, then for a given trajectory  $z_{[0,T+N-1]}$ of system $\Sigma$, we have that
\be\label{eq:Hz1}
    H_{L+N} (z_{[0,T+N-1]})\alpha = \bar{z}_{[0,L+N-1]}  
\ee 
for some $\alpha\in\Rbb^{T-L+1}$, as shown in Lemma~\ref{lem:persist}. In the same way, if the last statement holds, then, considering  $\mathcal{Z}_{[0,T-1]}$ where $\mathcal{Z} (k) = \bma z(k)^\top & \ldots& z(k+N)^\top \ema^\top $, we have that
\be \label{eq:Hzhat}
    H_L(\mathcal{Z}_{[0,T-1]})\alpha =  \bar{\mathcal{Z}}_{[0,L-1]}  
\ee 
also holds for the same $\alpha\in\Rbb^{T-L+1}$ as in \eqref{eq:Hz1}. Note that $\bar{\mathcal{Z}}_{[0,L-1]}$ is a rearranging and stacking of the elements in $\bar{z}_{[0,L+N-1]}$. Therefore the matrices $H_{L+N} (z_{[0,T+N-1]})$ and $H_L(\mathcal{Z}_{[0,T-1]})$ share the same rank.

By definition in the theorem, $U_\perp$ is a null space of ${{U} H_L(\mathcal{Z}_{[0,T-1]})}$, e.g., % Thus, we have that 
${{U} H_L(\mathcal{Z}_{[0,T-1]})} U_\perp=0$ holds. %, and, combining with the information given in 
This fact together with \eqref{eq:Hzhat} implies that %, we have that 
${U} H_L(\mathcal{Z}_{[0,T-1]})\alpha=0$, for any $\alpha\in\Rbb^{T-L+1}$ that satisfies \eqref{eq:Hzhat} with zero initial condition $\bar{z}_{[0,\nu-1]}=0$ (due to the only non-zero element of identity in $U$ as in \eqref{eq:U_def}). % for all trajectories $\bar{z}_{[0,L-1+N]}$. 
Therefore, from Finsler's lemma, the inequality in \eqref{eq:theo} is equivalent to
\[
   \alpha^\top H_{L}({\mathcal{Z}_{[0,T-1]}}  )^\top \Phi_L H_{L}({\mathcal{Z}_{[0,T-1]}}  )\alpha \geq 0,
\]
for all $\alpha$ as before (which results in admissible trajectories with zero initial conditions). Consequently, we have
%which results in 
\begin{equation} \label{eq:proof}
     \sum_{k=0}^{ L-1}\bar{\mathcal{Z}} (k)^\top\Phi_N\bar{\mathcal{Z}} (k) =
     \sum_{k=0}^{ L-1}\sum_{i,j=0}^N\bar{z}(k+i)^\top\Phi_{ij}\bar{z}(k+j) \geq 0.
\end{equation}
%If inequality \eqref{eq:proof} holds for any $\alpha\in\Rbb^{T-L+1}$ that satisfies~\eqref{eq:Hzhat}, then it is equivalent to say
Equivalently, we have 
\[ 
    \sum_{k=0}^{L-1}\bar{\mathcal{Z}}(k)^\top\Phi_N\bar{\mathcal{Z}}(k) = \bar{\mathcal{Z}}_{[0,L-1]}^\top\Phi_{L}\bar{\mathcal{Z}}_{[0,L-1]} \geq 0,
\]
for all trajectories $\bar{z}_{[0,L+N-1]}$ with initial conditions $\bar{z}_{[0,\nu-1]}=0$. By evaluating the $(L,\nu,N)$-$QSR$-dissipativity in \eqref{eq:LnuQSRdissip} and using the above inequality, it follows that  
%${\{\bar{z}(k)\}_{k=0}^{\nu-1}=0}$ 
% $\{\bar{z}(k)\}_{k=0}^{L-1{\color{red}+N}}$. 
%%%%%%%  $L-QSR$-dissipative The system \eqref{eq:sys_main} is $(L-v)QSR$-dissipative if, for a $\nu<L$, %the following holds
\be \label{eq:theo_L-QSR}
     \sum_{k=0}^{L-1-\nu}\tilde{\mathcal{Z}}(k+i)^\top\Phi_N\tilde{\mathcal{Z}}(k+j) = \sum_{k=0}^{L-1}\bar{\mathcal{Z}}(k+i)^\top\Phi_N\bar{\mathcal{Z}}(k+j)\geq 0 
\ee 
holds 
for any trajectory $\tilde{z}_{[0,L+N-1-\nu]}$ %$\{\tilde{z}(k)\}_{k=0}^{L-1{\color{red}+N}-\nu}$ 
with initial conditions $\tilde{x}_0=0$ and for any trajectory $\bar{z}_{[0,L+N-1]}$ %$\{\bar{z}(k)\}_{k=0}^{L-1{\color{red}+N}}$ 
with initial conditions $\bar{z}_{[0,\nu-1]=0}=0$. This proves the $(L,\nu,N)$-$QSR$-dissipativity of $\Sigma$. %${\{\bar{z}(k)\}_{k=0}^{\nu-1}=0}$.

Additionally, if \eqref{eq:theo_L-QSR} holds for any $n\leq \nu<L$ then the $(L,N)$-dissipativity follows immediately from the definitions in \eqref{eq:LQSRdissip} and \eqref{eq:LnuQSRdissip}. % it implies that the initial conditions of both $\tilde{x}$ and $\bar{x}$ are equal to zero, thus system~\eqref{eq:sys_main} is $L$-dissipative.
$\hfill \square$\\[0.2cm]

%%%%%%%%%%%%%%%%%%%%%%%%%%%%%%%%%%%%%%%%%%%%%%%%%%%%%%
%                       END PROOF
%%%%%%%%%%%%%%%%%%%%%%%%%%%%%%%%%%%%%%%%%%%%%%%%%%%%%%

Let us make a few remarks on the design choice of %Regarding the choices of parameters  
$\nu$, $L$ and $T$. From Theorem~\ref{theo:LQSR}, we can observe (from the way of using $U$ to constrain the admissible trajectories with zero initial conditions) that $\nu$ must be greater or equal to the order of the system in \eqref{eq:sys_main}, which is $n$.  
%we see that $\nu$ needs to be greater that the order $n$ of the system~\ref{eq:sys_main}, 
However, in practice, this information may not be obtained {\it a priori}. %it do know always have this information, therefore, 
As stated in \cite[Remark 1]{RBKA:19}, the parameter $\nu$ can be used as an upper bound of the order of the system. 
With regards to $L$, similar as before, the value must be greater than $\nu$ in order to have well-defined formulation in  Theorem~\ref{theo:LQSR}. Correspondingly, $L$ can be chosen arbitrarily larger than $\nu$.  %, we see that it needs to be greater than $\nu$, and can assume any value the designer desires. 
Finally, the choice of $T$  
%The reasoning of how to choose the value of $T$ 
is of utmost importance and comes from the requirement of $u$ being persistently excited of order $L+N+n$. In order to verify the latter, we can choose the value of $T$ such that the matrix $H_{L+N+n}(u_{[0,T+N-1]})$  is full row rank, that is $T\geq (L+n)(m+1)+Nm-1$.

Note that the requirement of $u$ being persistently exciting of order $L+N+n$ is only mentioned in the statement (i) of Theorem~\ref{theo:LQSR}. Therefore, the fulfilment of this condition and~\eqref{eq:theo} are not sufficient to guarantee that statement (ii) also holds. For the latter, as we mention in the proof of  Theorem~\ref{theo:LQSR} above, we construct $U_\perp=(UH_L(\mathcal{Z}_{[0,T+N-1]}))^\perp$ such that we generate all admissible trajectories $\bar{z}$ with zero initial conditions. For that, we have that the null space of $UH_L(\mathcal{Z}_{[0,T+N-1]})$ must exist for all $\nu$, $L$ and $T$. One way to guarantee that is by defining $T$ that depends on the dimension of %such feature is to define the size of the quantity $T$ looking at the size of 
$UH_L(\mathcal{Z}_{[0,T+N-1]})$, e.g.  %, that is, 
$T\geq\nu(m+p)+L-1$ should hold for all choices of $\nu$ and $L$. In order to remove the dependency on $\nu$, we can use the upper bound of the choices of $\nu$ where $n\leq\nu<L$. Therefore, if we choose $T$ such that $T\geq L(m+p+1)-1$, we guarantee that the null space of $UH_L(\mathcal{Z}_{[0,T+N-1]})$ exists for any choice of $L$ and $\nu$. {Correspondingly, if we verify Theorem~\ref{theo:LQSR} using some choice of $L$ and $\nu$, and $T$ such that the latter bound holds, then we guarantee the $L$-$QSR$-dissipativity without having to check the conditions for all~$\nu~(n\leq\nu<L)$. }

\section{Examples}

In this section, we present numerical simulations for checking the general dissipativity based on our data-driven test. For the setup,  
%For programming the conditions of Theorem~\ref{theo:LQSR}, 
we use the software Matlab (R2020a) with the help of a Windows 10 Enterprise LTSC computer, Intel Core i7-5600 (2.60 GHz), 16.0 GB RAM. 
%in conjunction with the parser YALMIP \cite{Lof} and the solver Mosek \cite{mos:19} for the optimization procedures of finding feasible solutions for Theorem~\ref{theo:LQSR} via linear matrix inequality (LMI) conditions. 

\subsection{$\mathcal{L}_2$-gain}
In this example we consider the case of searching for a minimal bound for the $\mathcal{L}_2$-gain of discrete-time systems with the aim to illustrate the applicability of the technique and also a comparison with the method described in \cite{RBKA:19}. 
With the help of the Matlab function \texttt{drss} and using  \texttt{rng(0)}, we randomly generate 200 stable systems of order 4 and outputs and inputs equal to 2. Using these systems, we generate trajectories of size $T_f=500$ admitting a normally distributed input with standard deviation of $10$ and mean zero and zero initial conditions of size 1.

We search for the minimal upper bound $\gamma$ for the $\mathcal{L}_2$-gain of the system such that $||y||_2 \leq \gamma^2 ||u||_2 $ holds for all trajectories ${z}_{[0,L-1]}$ with ${z}_{[0,\nu]}=0$. This is the equivalent to search for the minimal $\gamma$ such that the system is $(L,\nu,N)$-dissipative with respect to a $QSR$ supply function of the form~\eqref{eq:supplyfunction} with 
\[\Phi=\bma -I_p&0\\0 &\gamma^2 I_m \ema,\]
from which we see that $N=0$. 

For this example, we arbitrarily assume $L=30$ and $\nu=\{5,28\}$. 
We assume snapshots of size $T=(L+n)(m+1)-1 = 101$ for verifying Theorem~\ref{theo:LQSR} and Theorem~2 of \cite{RBKA:19}.
Note that by using the trajectories of size the requirements for the persistency of excitation of the input holds.
Additionally, this snapshot does not need to contain the initial conditions, it can be removed from any point of the trajectory. In our case, we assume this snapshot to start from point 50 of the generated trajectories. 
Regarding the search itself, we consider a simple bisection algorithm with a tolerance equal to $0.001$ and limit of iterations equal to 50. 
{ For applying Theorem~\ref{theo:LQSR} and Theorem~2 of \cite{RBKA:19}, we need to verify the non-negativeness of the matrix of the main inequality in both theorems. As recommended in \cite{RBKA:19}, we verify if the minimum real parts of the eigenvalues of the resulting matrices are positive or slightly negative, to which we allow a tolerance of $-1\times 10^{-8}$. }
Additionally, we present the optimal value of the $\mathcal{L}_2$-gain obtained using the command \texttt{norm(sys,inf)} of Matlab and the information on the models of the generated systems. 

\begin{figure}[h!]
	\begin{center}
  		\includegraphics[scale=0.6]{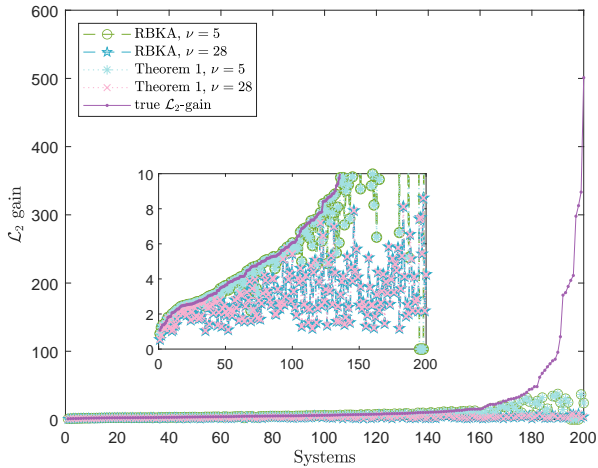}
		\caption{The plot of the estimated and theoretical $\mathcal{L}_2$-gain of $200$ stable systems using the methods in Theorem~\ref{theo:LQSR} and \cite[Theorem 2]{RBKA:19} (RBKA). }\label{fig:ex2RBKA1}
	\end{center}
\end{figure}

In Figure~\ref{fig:ex2RBKA1} we present the different values of $\gamma$ obtained using the methods describe in Theorem~\ref{theo:LQSR} and \cite[Theorem 2]{RBKA:19} (RBKA) with $\nu=\{5,28\}$. Additionally, we also present the theoretical $\mathcal{L}_2$-gain obtained using the model information for each system.
From this figure, we can see that the values obtained for the $\mathcal{L}_2$-gains using  { both Theorem~\ref{theo:LQSR} and \cite[Theorem 2]{RBKA:19}} for several systems are closer to the theoretical $\mathcal L_2$ gains, which, as mentioned in \cite{RBKA:19}, shows the applicability of the method for the estimation of the $\mathcal L_2$-gains. 
{ Note also, that the results obtained for Theorem~\ref{theo:LQSR} and \cite[Theorem 2]{RBKA:19} are the same, showing that, as expected, the conditions of Theorem~\ref{theo:LQSR} when using $N=0$ recover the conditions given in \cite[Theorem 2]{RBKA:19}.}
Additionally, note the difference on the results when applying the different values of $\nu$. When this value is low, the obtained gains are closer to the true values, however, for a higher $\nu$, the estimate becomes less accurate. {Additionally, observe that for two systems the method presented in this paper and the one from~\cite{RBKA:19} are } not able to find solutions for $\nu=5$, which could be caused by the complexity of the problem. { These two unfeasible cases are represented by zero in the figure.}

%%%%%%%%%%%%%%%%%%%%%%%%%%%%%%%%%%%%%%%%%%%%%%%%%%%%%%%%%%%%%%%%%%%%%%%%
%                             EXAMPLE 2 
%%%%%%%%%%%%%%%%%%%%%%%%%%%%%%%%%%%%%%%%%%%%%%%%%%%%%%%%%%%%%%%%%%%%%%%%
\subsection{Mass-spring-damper}

In this example we consider a typical mass-spring-damper system based on the results shown in Example 6.4 from~\cite{TW:97} and which is given by the following difference equation
\be \label{eq:sys_ex_msd}
    y(k+2) +  y(k+1) + y(k) = u(k).
\ee
For this mechanical system, we can construct explicitly the storage function. 
For that purpose, as in \cite{TW:97}, we consider the following storage function 
\begin{equation}\label{eq:V_discrete}
    V(k) = y^2(k+1) + y^2(k).
\end{equation}
%For showing dissipativity in the discrete-time setting, we need to show that 
%\[
%    V(k+1)-V(k) \leq w(y(k),y(k+1),\ldots,u(k),u(k+1),\ldots).
%\]
%Using the aforementioned choice of $V(k)$ in \eqref{eq:V_discrete}, 
It follows that
\begin{multline}
    V(k+1)-V(k)  %= y^2(k+2)+y^2(k+1)   \nonumber\\
%    & \qquad \qquad -y^2(k+1)-y^2(k)
%    \nonumber\\\nonumber
%    & = y^2(k+2) - y^2(k) \\\nonumber
%    & = \big(u(k)- y(k+1)-y(k)\big)^2 - y^2(k) \\\nonumber
    =  u^2(k)-2y(k)u(k) -2 y(k+1)u(k) \nonumber \\
     \qquad +   2 y(k)y(k+1) +y(k+1)^2 =: w_1(y(k),u(k)), \label{eq:ex_supply_msd}
\end{multline}
where the supply function $w_1$ takes the form of $1$-$QSR$ dissipative supply function as in \eqref{eq:supplyfunctionderi}  
%The right side of the last equation is the supply rate that we are looking for, which we call by $w_1(y(k),u(k))$ and we can write it in the form of \eqref{eq:supplyfunctionderi} 
with %the following choices
\[ 
    \Phi_{00} = \bma 0 & -1 \\ -1 & 1\ema, ~ 
    \Phi_{10} = \bma 1 & -1 \\ 0 & 0\ema,~
    \Phi_{11} = \bma  1 & 0 \\ 0 & 0\ema.
    \]
%which shows that this system is $QSR$-dissipative with respect to a general supply rate. 

Given this prior knowledge on the system, we perform two main tests. The first one is taken to show that the system is indeed not $QSR$ dissipative with a common supply rate as in \cite{RBKA:19}, but it is instead $1$-$QSR$ dissipative with respect to a general supply rate $w_1$. The second test is performed to show that the system is $(L,1)$-$QSR$-dissipative in the sense of Theorem \ref{theo:LQSR}.

In order to perform the tests, we generate 1000 different samples each with 300 discrete-time points (e.g., $T_f=300$), where we consider zero initial conditions of size $2$ for both the output and input signals. We assume a normally distributed input with a standard deviation of $10$ and mean zero.

\begin{figure}[h!]
	\begin{center}
  		\includegraphics[scale=0.6]{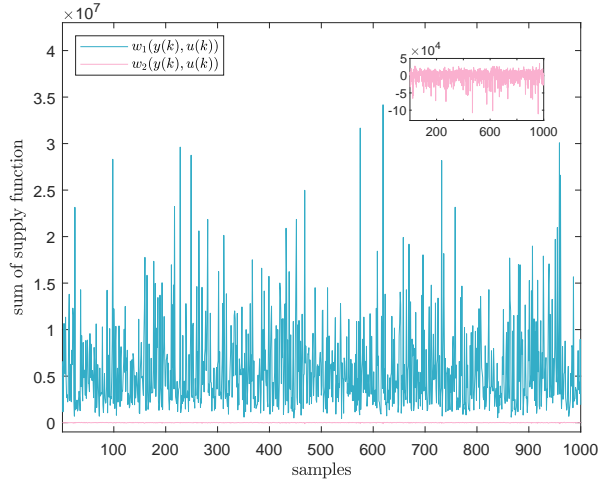}
		\caption{Comparison between the different supply functions }\label{fig:comparison_supplyfunc}
	\end{center}
\end{figure}

As mentioned before, in the first part of this example we want to verify if the system is $QSR$ dissipative with respect to the general supply rate $w_1$ and to the supply rate of the form of $w_1$ with only the terms depending on $k$, that is, 
\[ 
    w_2(y(k),u(k)) =  u^2(k)-2y(k)u(k).
\]
Thus, we verify if 
\be\label{eq:ex_supplyTf}
    \sum_{k=0}^{T_f-1-N} w(u(k),y(k)) \geq0,
\ee
with $w_1$ and $w_2$, holds for all the obtained trajectories. 
In Figure~\ref{fig:comparison_supplyfunc}, we can observe the values of the left side of~\eqref{eq:ex_supplyTf} obtained for each sample considering the general supply function $w_1$ and $w_2$.

As one can see in Figure~\ref{fig:comparison_supplyfunc}, we have that \eqref{eq:ex_supplyTf} holds considering the general supply function $w_1$ for all the samples tested, while considering the supply rate in $w_2$, we cannot guarantee that \eqref{eq:QSRdissip} always holds. This shows that indeed the system is not dissipative when considering only the terms depending on $k$ in the supply function, but instead, it is dissipative with respect to a general quadratic supply function. 

For the second part of this example, we verify the $(L,1)$-$QSR$-dissipativity of system~\eqref{eq:sys_ex_msd} with different values of $\nu$, $L$ and $T$. 
Taking, for instance, $L=(n+1,\ldots,10)$, we can test whether Theorem~\ref{theo:LQSR} holds for all or some $n\leq\nu<L$ considering snapshots of length $T$ from all 1000 trajectories.
As for the previous example, we assume this snapshot to start from point 50 of the trajectories. 
Also, we consider two cases for the choice of $T$, one with respect to condition of the input being persistently exciting  ($T_1=(L+n)(m+1)+Nm-1$ with $N=1$) and another considering that we guarantee that $UH_{L}(\mathcal{Z}_{[0,T-1]})$ has a null space ($T_2 = (m+p+1)L-1$). 
Using such choices, we have that Theorem~\ref{theo:LQSR} holds for all choices of $L$, $\nu$ and $T$. Therefore, using such choices of parameters we can show that, for the values of $L$ tested, the system is $(L,1)$-$QSR$-dissipative with respect to the supply function~$w_1$, which is already expected given that the system is dissipative with respect to the supply function~$w_1$. Note that considering $T_1$, we are able to obtain  $U_\perp=(UH_{L}(\mathcal{Z}_{[0,T-1]}))^\perp$ for all choices of $L$ and $\nu$, showing that for this system the choice of $T$ based only on the persistency of excitation of $u$ condition is sufficient to test the dissipativity of the system using Theorem~\ref{theo:LQSR}. However, for different systems, verifying only this condition can lead to not enough data such that we do not obtain~$U_\perp$.

\section{CONCLUSIONS}

We proposed a method to verify the dissipativity of discrete-time LTI systems with respect to a quadratic general QSR supply function using only one shot of data. With this new formulation, we are able to verify the general dissipativity that applies to any LTI systems as studied before in the context of behavioural framework \cite{TW:97}.

%deal with systems where it is required or at least desired to take the time-differences in consideration in the process of designing the supply function, which is the case, for instance, of time-delay systems. 
%
We presented two examples to show the potential of our method, and also to illustrate the reasoning for the choices of the required parameters to solve our main results. %The identification of specific form of a general quadratic supply-rate function using our approach has several potential applications., such as stability analysis of interconnected general LTI systems, design of  control-by-interconnection and the analysis of interconnected model-based and data-driven systems. 
%
%This approach has several possible extensions, for instance, the interconnection of data-driven and model-based systems through the analysis of the interconnection of their supply function; the dissipativity of systems affected  by noise and perhaps with uncertainties; and, the stability analysis from the perspective of verifying the time-difference of a Lyapunov function in the same way as we analyse the storage function. 

\addtolength{\textheight}{-12cm}   % This command serves to balance the column lengths

%%%%%%%%%%%%%%%%%%%%%%%%%%%%%%%%%%%%%%%%%%%%%%%%%%%%%%%%%%%%%%%%%%%%%%%%%%%%%%%%

%%%%%%%%%%%%%%%%%%%%%%%%%%%%%%%%%%%%%%%%%%%%%%%%%%%%%%%%%%%%%%%%%%%%%%%%%%%%%%%%

%%%%%%%%%%%%%%%%%%%%%%%%%%%%%%%%%%%%%%%%%%%%%%%%%%%%%%%%%%%%%%%%%%%%%%%%%%%%%%%%
%\section*{APPENDIX}


%\section*{ACKNOWLEDGMENT}

%The preferred spelling of the word ÒacknowledgmentÓ in America is without an ÒeÓ after the ÒgÓ. Avoid the stilted expression, ÒOne of us (R. B. G.) thanks . . .Ó  Instead, try ÒR. B. G. thanksÓ. Put sponsor acknowledgments in the unnumbered footnote on the first page.

\bibliographystyle{ieeetr}
\bibliographystyle{alpha}
\bibliography{references}

\begin{thebibliography}{10}

\bibitem{Doyle1989}
J.~C. Doyle, K.~Glover, P.~P. Khargonekar, and B.~A. Francis, ``State-space
  solutions to standard $\mathcal{H}_2$ and $\mathcal{H}_\infty$ control
  problems,'' {\em IEEE Transactions on Automatic Control}, vol.~34, no.~8,
  pp.~831--847, 1989.

\bibitem{Wil:72}
J.~C. Willems, ``Dissipative dynamical systems part {I}: General theory,'' {\em
  Archive for rational mechanics and analysis}, vol.~45, no.~5, pp.~321--351,
  1972.

\bibitem{Hill1976}
D.~J. Hill and P.~J. Moylan, ``The stability of nonlinear dissipative
  systems,'' {\em IEEE Transactions on Automatic Control}, vol.~21, no.~5,
  pp.~708--711, 1976.

\bibitem{HM:80}
D.~J. Hill and P.~J. Moylan, ``Dissipative dynamical systems: Basic
  input-output and state properties,'' {\em Journal of the Franklin Institute},
  vol.~309, no.~5, pp.~327--357, 1980.

\bibitem{Lanzon2008}
A.~Lanzon and I.~R. Petersen, ``{Stability Robustness of a Feedback
  Interconnection of Systems With Negative Imaginary Frequency Response},''
  {\em IEEE Transactions on Automatic Control}, vol.~53, no.~4, pp.~1042--1046,
  2008.

\bibitem{Ouyang2014}
R.~Ouyang and B.~Jayawardhana, ``{Absolute stability analysis of linear systems
  with Duhem hysteresis operator},'' {\em Automatica}, vol.~50, no.~7,
  pp.~1860--1866, 2014.

\bibitem{Ouyang2013}
R.~Ouyang, V.~Andrieu, and B.~Jayawardhana, ``{On the characterization of the
  Duhem hysteresis operator with clockwise input-output dynamics},'' {\em
  Systems \& Control Letters}, vol.~62, no.~3, pp.~286--293, 2013.

\bibitem{TW:97}
H.~Trentelman and J.~Willems, ``Every storage function is a state function,''
  {\em Systems \& Control Letters}, vol.~32, no.~5, pp.~249--259, 1997.

\bibitem{WRMDM:05}
J.~C. Willems, P.~Rapisarda, I.~Markovsky, and B.~L. {De Moor}, ``A note on
  persistency of excitation,'' {\em Systems \& Control Letters}, vol.~54,
  no.~4, pp.~325--329, 2005.

\bibitem{dPT:19}
C.~De~Persis and P.~Tesi, ``Formulas for data-driven control: Stabilization,
  optimality, and robustness,'' {\em IEEE Transactions on Automatic Control},
  vol.~65, no.~3, pp.~909--924, 2019.

\bibitem{BA:19}
J.~Berberich and F.~Allg{\"o}wer, ``A trajectory-based framework for
  data-driven system analysis and control,'' {\em arXiv preprint
  arXiv:1903.10723}, 2019.

\bibitem{MR:08}
I.~Markovsky and P.~Rapisarda, ``Data-driven simulation and control,'' {\em
  International Journal of Control}, vol.~81, no.~12, pp.~1946--1959, 2008.

\bibitem{MM-MR:17}
T.~Maupong, J.~C. Mayo-Maldonado, and P.~Rapisarda, ``On lyapunov functions and
  data-driven dissipativity,'' {\em IFAC-PapersOnLine}, vol.~50, no.~1,
  pp.~7783--7788, 2017.

\bibitem{RBKA:19}
A.~Romer, J.~Berberich, J.~K{\"o}hler, and F.~Allg{\"o}wer, ``One-shot
  verification of dissipativity properties from input--output data,'' {\em IEEE
  Control Systems Letters}, vol.~3, no.~3, pp.~709--714, 2019.

\bibitem{vWdPCT:20}
H.~J. van Waarde, C.~De~Persis, M.~K. Camlibel, and P.~Tesi, ``Willems’
  fundamental lemma for state-space systems and its extension to multiple
  datasets,'' {\em IEEE Control Systems Letters}, vol.~4, no.~3, pp.~602--607,
  2020.

\bibitem{dOS:01}
M.~C. de~Oliveira and R.~E. Skelton, ``Stability tests for constrained linear
  systems,'' in {\em Perspectives in robust control}, pp.~241--257, Springer,
  2001.

\bibitem{BLME:07}
B.~Brogliato, R.~Lozano, B.~Maschke, and O.~Egeland, {\em Dissipative systems
  analysis and control: Theory and Applications}, vol.~2.
\newblock Springer, 2007.

\bibitem{KA:10}
N.~Kottenstette and P.~J. Antsaklis, ``Relationships between positive real,
  passive dissipative, \& positive systems,'' in {\em Proceedings of the 2010
  American control conference}, pp.~409--416, IEEE, 2010.

\end{thebibliography}

\end{document}